\documentclass{scspaperproc}

\usepackage{latexsym}
\usepackage{graphicx}
\usepackage{mathptmx}

\usepackage{amsmath}
\usepackage{amsfonts}
\usepackage{amssymb}
\usepackage{amsbsy}
\usepackage{amsthm}

\usepackage[pdftex,colorlinks=true,urlcolor=blue,citecolor=black,anchorcolor=black,linkcolor=black,bookmarks=false]{hyperref}

\usepackage{hyphenat}
\newtheoremstyle{scsthe}
{8pt}
{8pt}
{\it}
{}
{\bf}
{.}
{.5em}
{}

\theoremstyle{scsthe}

\begin{document}

%
%
\SCSpagesetup{Salazar-Serna, Cadavid, and Franco}

\def\SCSconferencename{Annual Simulation Conference}

\def\SCSconferenceacro{ANNSIM'24}

\def\SCSpublicationyear{2024}

\def\SCSconferenceeditors{P.J. Giabbanelli, I. David, C. Ruiz-Martin, B. Oakes and R. C\'{a}rdenas}

\def\SCSconferencedates{May 20-23}

\def\SCSconferencevenue{American University, DC, USA}

\title{ANALYZING TRANSPORT POLICIES IN DEVELOPING COUNTRIES WITH ABM}

\author[\authorrefmark{1} \authorrefmark{2}]{Kathleen Salazar-Serna}
\author[\authorrefmark{2}]{Lorena Cadavid}
\author[\authorrefmark{2}]{Carlos J. Franco}

\affil[\authorrefmark{1}]{Department of Civil en Industrial Engineering, Pontificia Universidad Javeriana - Cali, Colombia}
\affil[ ]{\textit {kathleen.salazar@javerianacali.edu.co}}

\affil[\authorrefmark{2}]{Department of Computer and Decision Sciences, Universidad Nacional de Colombia - Medellín, Colombia}
\affil[ ]{\textit{dlcadavi@unal.edu.co, cjfranco@unal.edu.co}}

\maketitle

\section*{Abstract}

Deciphering travel behavior and mode choices is a critical aspect of effective urban transportation system management, particularly in developing countries where unique socio-economic and cultural conditions complicate decision-making. Agent-based simulations offer a valuable tool for modeling transportation systems, enabling a nuanced understanding and policy impact evaluation. This work aims to shed light on the effects of transport policies and analyzing travel behavior by simulating agents making mode choices for their daily commutes. Agents gather information from the environment and their social network to assess the optimal transport option based on personal satisfaction criteria. Our findings, stemming from simulating a free-fare policy for public transit in a developing-country city, reveal significant influence on decision-making, fostering public service use while positively influencing pollution levels, accident rates, and travel speed.

\textbf{Keywords:} bus rapid transit, transport modes, travel behavior, free-fare, urban mobility.

\section{Introduction}
\label{sec:intro}

The analysis of travel mode choice holds substantial relevance in urban planning, particularly for devising strategies to enhance citizens' quality of life, promote public transport usage, and facilitate sustainable urban development. With the rapid growth of cities and escalating transport demands, effective policies are crucial to establish access to efficient and sustainable transportation systems. Travel mode choice has been analyzed to address a variety of issues, for instance, traffic congestion \cite{ali2021travel}, increase of private vehicles ownership \cite{kangur2017agent}, active mode choices promotion \cite{ali2022identification}, shared mobility, ride-hailing services, among others \cite{hawas2016multi}. Traditionally, discrete choice models like multinomial logit and multinomial probit have been widely accepted for studying travel mode choice \cite{williams1977formation}, but recent research advocates for more dynamic approaches. While various studies have traditionally employed statistical analyses to investigate travel mode choice in urban contexts, discrete choice models struggle to capture the nonlinear relationships and constant interactions in complex transportation systems. Agent-based simulation models (ABM) have emerged as a valuable alternative, offering the flexibility to represent dynamic interactions in such systems \cite{kagho2020agent}.

A few works have applied ABM to analyze travel mode choices. For instance, Faboya et al. \cite{faboya2020using} analyzes the adaptive travel behavior of individuals commuting to and from a university. In a different study Kangur et al.\cite{kangur2017agent} explores consumer behavior in adoption of electric vehicles. Another example is Chen and Cheng \cite{chen2010review} research which simulates decision making behaviors under different sets of traffic conditions. Although several works have utilized ABM to analyze travel mode choices, the majority have focused on developed countries. This is a notable gap, as urban mobility in developing countries, particularly in the Global South, is distinct, with private vehicles, especially motorcycles, dominating due to inadequate public transport coverage and reliability \cite{Hagen2016}. The prevalence of motorcycles, primarily used by middle/low-income individuals for commuting \cite{eccarius2020adoption}, contributes to externalities like traffic congestion, air pollution, and road accidents \cite{suatmadi2019demand}. Addressing these challenges requires policymakers in developing countries to devise sustainable transportation policies aligned with the goals set for 2030.      

This research presents the outcomes of an agent-based simulation model that captures the interactions among urban travelers as they decide on their primary mode of transport. Notably, this is the first agent-based model tailored to analyze commuter behavior in developing countries, marking a pioneering effort that also includes motorcycles as a mode of transportation. To ensure realistic human behavior representation within the chosen context, a Colombian city is selected as the case study. The objective of this paper is to illustrate the potential impact of implementing a fare-free policy for public transit, aimed at mitigating the rise in private vehicle usage. Despite gaining significant attention in 2023, especially after being mentioned by the Colombian president on social media \cite{papernews}, this strategy has yet to be implemented in any city in the country, and no studies have been conducted to date.

The remainder of the paper presents a brief description of the simulation model in section 2. Section 3 describes the experiments and results are discussed in the following section. Finally, section 4 covers conclusions and future work.   

\section{Simulation model}

The model simulates the travel mode choices made by urban commuters. Agents are created with demographic attributes such as sex, age, and income level and then distributed in neighborhoods throughout the city according to their socioeconomic status. To represent social influence in decision-making, agents are connected through a social network using a scale-free topology. Preferential attachment is a widely used model for large real-world networks exhibiting a power-law distribution \cite{janssen2003simulating}. In this model, agents connect to other nodes chosen at random, with a bias towards nodes already highly connected. People within the same socioeconomic group have a higher probability of being connected. This represents the concept of forming social ties with similar others \cite{carley1991theory}.In a previous work related to this research, we explored the impact of using different types of network and explained the reasons for using a preferential attachment network \cite{salazar2023social}. In the supplementary material, an example of a social network simulated with our model can be found. The simulation begins with commuters traveling by motorcycle, car, or public transit at the peak-hour, with the goal of maximizing satisfaction on their journey to the designated workplace. After completing the trip, they assess travel satisfaction using a utility function that incorporates various transport attributes, as detailed in Equation \eqref{eq:satisf}. 

\begin{equation} \label{eq:satisf}
S = \sum_{i=1}^nV_i * W_i.
\end{equation}

Where: S is the overall satisfaction of the agent. It results from the weighted sum of the values \mbox{$V_i$} obtained for the transport attributes after the trip and their associated weights \mbox{$W_i$}. The weights are determined based on the socioeconomic and cultural setting of the users and are understood as the level of importance given to the respective transport attribute. In our model, they were parameterized by socioeconomic level, according to the results of 970 responses to a survey we conducted in the city selected as a case study \cite{survey}. The function considers the following attributes: acquisition cost, operating cost, road safety, personal security, comfort of travel, commute time, and emitted pollution. Values for these attributes are affected by the state of the system as a result of individual and aggregate decisions by agents. Aspects such as congestion, road accidents, and CO2 emissions vary depending on the number of cars, motorcycles, or buses in circulation. Each agent calculates a level of density of vehicles in each time step, counting the number of equivalent cars in the surrounding patches. The density level affects the travel speed with a logarithmic decrease, and ultimately the kilometers traveled, the travel time, and the emissions per kilometer change accordingly.           

Based on the CONSUMAT approach \cite{jager2012updated}, which is a framework that integrates different theories of consumer behavior, we developed a mental model module that helps agents make a decision about the transport mode. By comparing the satisfaction obtained to a satisfaction threshold and simultaneously comparing the uncertainty about the satisfaction that will be obtained with the selected mode to an uncertainty threshold, agents implement one of these strategies: repeat, imitate, deliberate, or inquire. The uncertainty is calculated as \mbox{$U = \alpha * (1- \% times-using-mode) + (1 - \alpha) * (1 - \% peers-using-mode)$}; it is a combination of their personal user experience and the experiences of their similar others (peers within their social network). As an extension of the CONSUMAT approach, we introduce this calculation of uncertainty that incorporates social influence in decision making, taking into account that people gather information with their peers when making decisions under uncertainty \cite{janssen2003simulating}. Alpha is a parameter between zero and one that balances the weight that individuals give to the personal and collective experience; even if the commuter has little previous user experience with the transport mode, she can still obtain information from neighbors on the social network, which might help reduce uncertainty. For our model, alpha was fixed at 0.48, according to Hofstede's measurement of cultural dimensions: acceptance of uncertainty and individualism for Colombia. It is identified as a country with a very collectivist society that gives high value to the group's opinion and acceptance. This concept is explained in more detail in Sections 2.2 and 3.1 of Salazar-Serna et al. \cite{salazar2023simulating}. Note that the uncertainty level leads to different strategies to make decisions, depending on whether this is below or above the uncertainty threshold. The model was implemented with NetLogo 6.4.0, in which time steps are counted in ticks. One tick corresponds to two minutes of a peak hour of a typical year in the real world. The decision period for agents to make a decision about continuing using the same mode or change is one year.

The validation of the model was carried out using the 'validation in parts' technique \cite{carley1996validating}. It suggest to validate separately inputs, processes, and outputs. Inputs were validated to follow the distribution in a real system. A conceptual model based on travel mode choice literature was developed and experts in simulation were consulted to validate its logic. The code procedures were incrementally validated by modules and consistency was checked with extreme values. Outputs of 100 runs with a tailored scenario were compared to real world patterns for the case study city and emerging behaviors matched with those from the historical data (supplementary material presents plots for this comparison).

\section{Virtual Experiments}
Empirical data were collected in a Colombian city, serving as a reference to inform the model and resemble a city in a developing country. Taking into account the performance limitation of NetLogo with a large number of agents \cite{netlogo}, simulations were run with different scales to check at which level the social network meets the properties of a preferential attachment network. A synthetic society was created using a scale of 1:1000 to represent 1.2 million transport users. Satisfaction and uncertainty thresholds are normally distributed by transport mode for each individual. Independent variables such as the initial distribution of transport users, agents' demographics, technical characteristics of travel modes, and road accident rates are initialized according to values reported in the supplementary material. The outputs of interest are users' distribution at the end of each decision period, average travel speed, generated pollution, and accident rates. The analysis period is 10 years. Agents evaluate information annually to make decisions regarding mode changes. The simulation is run 100 times, and the average results are calculated at each decision period. Indicators are calculated for a base case scenario compared to a hypothetical scenario of implementing fare-free public transit.         

\section{Results}
\begin{figure}[htb]
{
\centering
\includegraphics[width=1\textwidth]{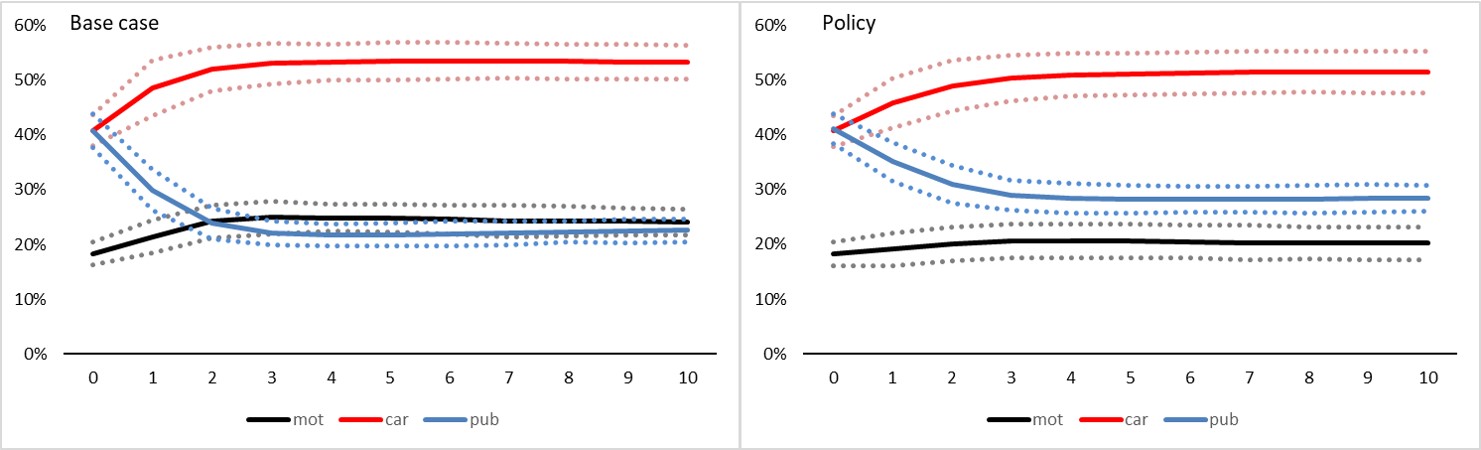}
\caption{Proportion of transport users by mode.Dotted lines represent C.I. 95\%}\label{fig:use}
}
\end{figure}

This section presents results obtained with an agent-based simulation model that represents travel mode choices of urban commuters in developing countries and analyzes the impacts on the system of implementing transport policies. This paper compares a base-case scenario with the implementation of a fare-free policy for public transit. Figure \ref{fig:use} shows the results for the distribution of users before and after the policy. Transport users in the base case migrate from public transit to private options in a great proportion. Without changes introduced in the system, the number of cars will increase by around 34\%, which means approximately 175,000 more cars in circulation in the next 10 years. Dissatisfaction with the public service leads to a drop in the number of users, falling from 41\% to 23\%, which is consistent with current trends in the real system \cite{magazinenews} due to the low quality of the service and the high insecurity rates (see the supplementary material). The popularity of motorcycles increases slightly over the years, yet an additional 82,000 units could be expected to be registered over the next decade.      

\begin{figure}[ht]
     \centering
     \subcaptionbox{Accident rate.\label{subfig:a_acc}}{
         \includegraphics[width=0.32\textwidth]{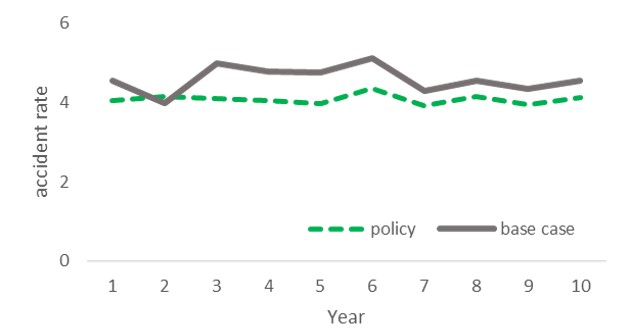}
     }
     \subcaptionbox{Pollution.\label{subfig:b_poll}}{
         \includegraphics[width=0.29\textwidth]{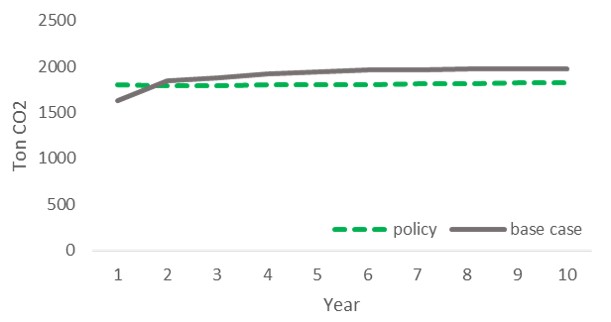}
     }
          \subcaptionbox{Travel speed.\label{subfig:c_speed}}{
         \includegraphics[width=0.29\textwidth]{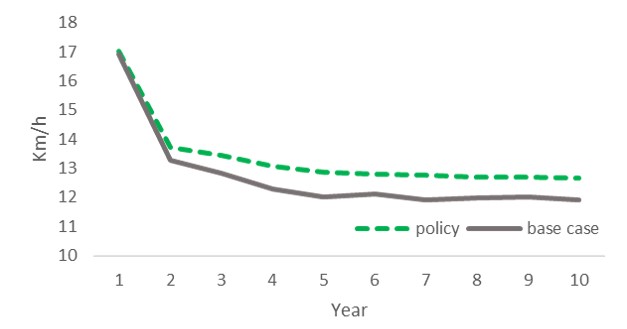}
     }
     \caption{Results before and after policy implementation.}
     \label{fig:indicators}
\end{figure}

The distribution of transport users changes with the implementation of the policy. Participation in public transit drops to 29\%, keeping a higher percentage of users than in the base case and avoiding a greater increase in private vehicle ownership. This is explained by higher levels of satisfaction with the transport mode, especially among users with low income levels who give great importance to the acquisition and operating cost of the transport mode. Those migrating to motorcycles - the most affordable option for them in the base case scenario - stay in the public service when the fare cost is zero. Also, the increase in cars is lower with the policy implementation; still, some users with greater economic power who value comfort and personal security more than those in lower socioeconomic classes continue to choose the car. In this scenario, an increase of up to 30\% is expected for cars and 14\% for motorcycles. Figure \ref{fig:indicators} presents three key indicators that show the beneficial impact of the policy on critical issues such as road accidents, pollution, and congestion. The average accident rate (number of accidents per 100,000 people) is lower when more users use public transport. Motorcycles are the transport mode that makes the highest contribution to the consolidated rate, with rates above 10 per 100,000 people (see supplementary material).           

Pollution is measured in tons of CO2. In the base scenario, there are more cars and motorcycles in circulation, increasing the CO2 emissions (see supplementary material). Implementing the policy could contribute by reducing 72 tons during commute at peak hour. The average speed in the system during rush hour is expected to decrease as the number of cars and motorcycles increases over the years in the base case. On the other hand, with more users in public transit, the average speed is steady; however, this is the mode with the lowest average speed, while the motorcycle is the fastest.

\section{Conclusions}
This paper presents an impact evaluation of a public transport policy that offers free transit to urban commuters in developing economies with predominant ownership of private vehicles. Using census and survey data from a Colombian case study city, we conducted experiments using an agent-based simulation model to represent the decision-making process for travel-mode choices. The findings indicate that the implementation of the policy influences the use of public transport. This is attributed to high satisfaction among users for whom cost is the most valued transport attribute. Consequently, reduced migration from public to private alternatives positively impacts the city by lowering accident rates and CO2 emissions, critical issues in large cities with respiratory complications due to air pollution and significant traffic accident fatalities. Moreover, the policy contributes to an improved average travel speed during peak hours, due to a reduced number of cars and motorcycles in circulation compared to the base-case scenario. If no policy or change is introduced into the system, private vehicle ownership is expected to increase in subsequent years, exacerbating the externalities of their extensive use. The fare-free policy counteracts the decrease of public transport usage; nevertheless, the proportion of users continues to decrease due to other aspects considered in the selection of mode, such as insecurity, which is actually a major concern for passengers in the city. Although it is not within the scope of this work, further analysis is required to evaluate the cost effectiveness of the fee-free policy, comparing it with the social positive impacts obtained. For future work, we intend to analyze mode-shift dynamics within specific demographic groups, facilitating the identification of intervention factors for targeted public policies. In addition, a second case study is being analyzed to compare the results, and future work will test a case in a different country. This model serves as a test-bed that can be parameterized for different territories considering their economic, social, and cultural conditions. We hope that this research becomes a valuable reference for policymakers and contributes to the development of improved transport systems especially in developing countries.

\section{Supplementary material}
Found additional information in: \url{https://github.com/Kathleenss/ANNSIM2024-Supplementary-material}

\bibliographystyle{scsproc}

\bibliography{Ref}


\section*{Author Biographies}

\textbf{\uppercase{Kathleen Salazar-Serna}} is a PhD candidate in the Department of Computer and Decision Sciences at Universidad Nacional de Colombia - Medellín and an assistant professor at the School of Engineering and Sciences at Pontificia Universidad Javeriana in Cali. Her current research interests focus on sustainability issues and transport policy analysis. She uses agent-based modeling and network analysis to study transport dynamics. Her email address is \email{kathleen.salazar@javerianacali.edu.co}.

\textbf{\uppercase{Lorena Cadavid}} is a professor in the Department of Computer and Decision Sciences at the Universidad Nacional de Colombia - Medellín. In addition to her academic role, she is an enterprise consultant who applies her expertise to guide organizations towards data-driven decision making. Her research interest lies in policy design through modeling and simulation of social phenomena and uses data analysis to support entrepreneurial decision making. Her email is   
\email{dlcadavi@unal.edu.co}.

\textbf{\uppercase{Carlos J. Franco}} works as a full professor in the Department of Computer and Decision Sciences at Universidad Nacional de Colombia - Medellín. His research areas include complex systems, energy market modeling and simulation, and policy evaluation and strategy formulation. His email is   
\email{cjfranco@unal.edu.co}.

\end{document}


%
%
\SCSpagesetup{Salazar-Serna, Cadavid, and Franco}

\def\SCSconferencename{Annual Simulation Conference}

\def\SCSconferenceacro{ANNSIM'24}

\def\SCSpublicationyear{2024}

\def\SCSconferenceeditors{P.J. Giabbanelli, I. David, C. Ruiz-Martin, B. Oakes and R. C\'{a}rdenas}

\def\SCSconferencedates{May 20-23}

\def\SCSconferencevenue{American University, DC, USA}

\title{ANALYZING TRANSPORT POLICIES IN DEVELOPING COUNTRIES WITH ABM}

\author[\authorrefmark{1} \authorrefmark{2}]{Kathleen Salazar-Serna}
\author[\authorrefmark{2}]{Lorena Cadavid}
\author[\authorrefmark{2}]{Carlos J. Franco}

\affil[\authorrefmark{1}]{Department of Civil en Industrial Engineering, Pontificia Universidad Javeriana - Cali, Colombia}
\affil[ ]{\textit {kathleen.salazar@javerianacali.edu.co}}

\affil[\authorrefmark{2}]{Department of Computer and Decision Sciences, Universidad Nacional de Colombia - Medellín, Colombia}
\affil[ ]{\textit{dlcadavi@unal.edu.co, cjfranco@unal.edu.co}}

\maketitle

\section*{SUPPLEMENTARY MATERIAL}

This document contains additional information to supplement the paper presented at the ANNSIM 2024 conference. The first section provides details on the social network, and the next section includes a diagram with the decision-making process. Section C contains details about the validation of the model using historical data and the comparison with the diffusion curve according to the model of Bass,    and the following section indicates an external link that contains results from the survey applied to transport users in our case study city and serves as a source to parametrize some of the variables in the model. The final section redirects to a working paper that presents the analysis of the survey results. 

\appendix

\section{Social Network} \label{app:satisfaction}

\begin{figure}[htb]
{
\centering
\includegraphics[width=0.5\textwidth]{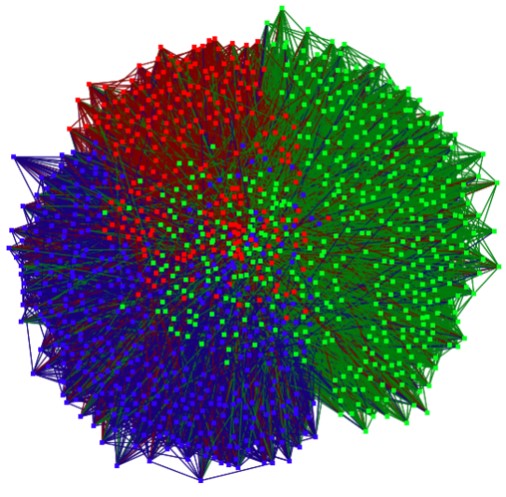}
\caption{Simulated social network with 1,250 synthetic transport users. Network matrix generated in NetLogo 6.4. Graph processed in ORA software. Blue agents belong to low-class status. Red corresponds to high-class people. Green represents agents in middle-class.}\label{fig:net}
}
\end{figure}

\section{Decision-making module} \label{app:flowchart}
Agents make decisions following the CONSUMAT approach. Depending on the comparison of the satisfaction and uncertainty levels with the thresholds, agents follow one of these strategies: repeat, imitate, deliberate, or inquiry.

\begin{figure}[htb]
{
\centering
\includegraphics[width=0.5\textwidth]{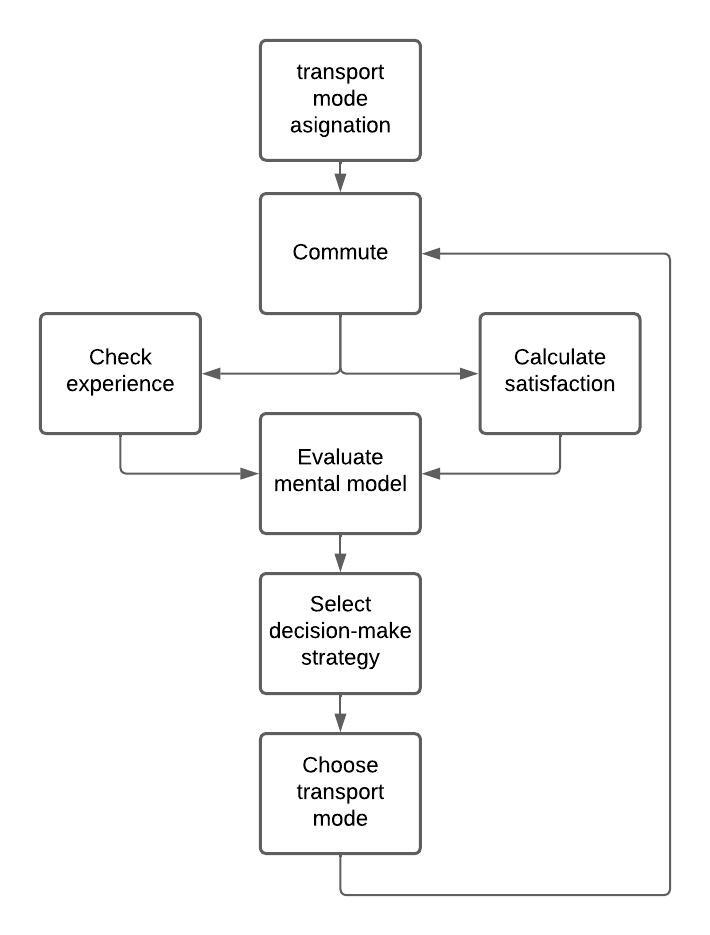}
\caption{Flowchart of the decision-making process for agents.}\label{fig:flow}
}
\end{figure}

\section{Comparison of simulation results versus historical data } \label{app:validation}   
Figure 3 presents the results for the average transport users of 100 runs with the simulation model. The simulation was initialized using the real distribution of users in 2018 and the sociodemographic attributes of the agents were parameterized using information from Cali city. We compared the simulated results with the real percentage of users before the pandemic. It can be observed that the simulation represents the general patterns of the real system, having an increase for private options and a decrease of public transit users. The average percentage in the third period matches those percentages in historical data in the year 2020 (period 2 in the plots) \cite{cali}.  

To have a point of reference for the forecast of the simulation, we contrast plots of the S-curve calculated with the Bass model for motorcycles (See Figure 4). We present the real data between 2007 and 2023 and the subsequent years present the forecast with both the nonlinear reference of Bass and the simulation.   

\begin{figure}[htb]
{
\centering
\includegraphics[width=1\textwidth]{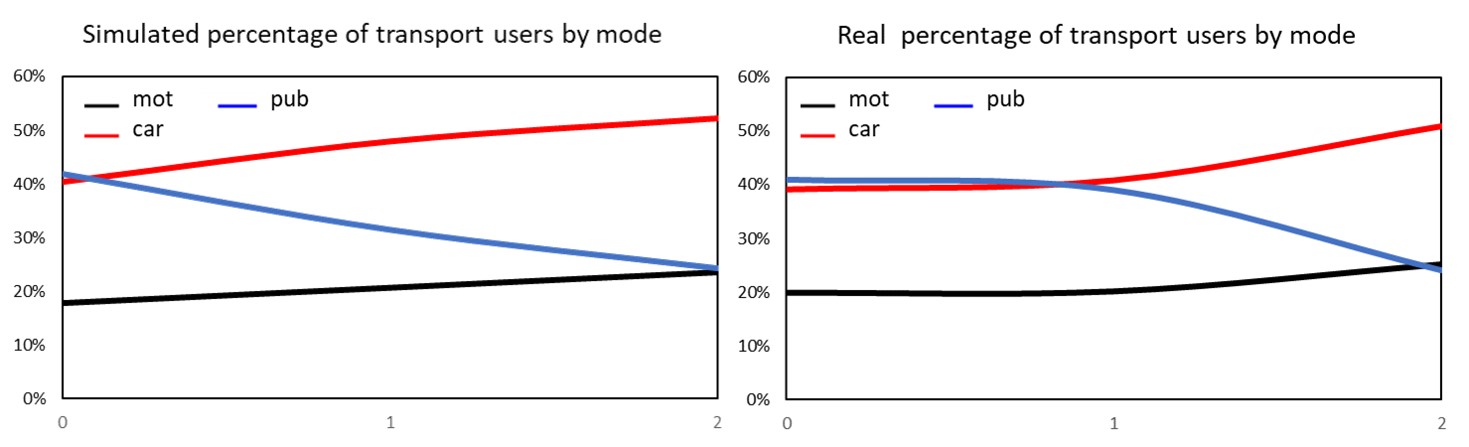}
\caption{Average percentage of transport users by mode. Simulated versus real data plots}\label{fig:valid}
}
\end{figure}

\section{survey results}\label{app:survey}
A summary with results of a survey conducted in the city of the case study that was used to determine and calibrate the parameters of the model is available in this link: \url{https://public.tableau.com/app/profile/jes.s.d.az.blanco/viz/Encuestas_16844425109290/Surveysummary}

\section{Experiment parameters}\label{app:parameters}
Table \ref{tab:param} presents some of the most important parameters to initialize the simulation, differentiated by attributes of the transport mode (Table 1) and the parameters that allow the customization of the model for a specific city (Table 2). 

\begin{table}[htb]
\caption{Experiment parameters associated with transport modes.}\label{tab:param}
\centering
\begin{tabular}{ll}
\hline
Parameter & Value\\ \hline
\% emissions motorcycle & 126g/km CO2 \\
\% emissions car & 204g/km CO2\\
efficiency motorcycle & 120km/gal\\
efficiency car & 50km/gal\\
\hline
\end{tabular}
\end{table}

\begin{table}[htb]
\caption{Parameters associated with the city.}\label{tab:param}
\centering
\begin{tabular}{llll}
\hline
Parameter & Value & Parameter & Value\\ \hline
\% income-level 1 & 34\% & average speed motorcycle -peak hour & 20km/h\\
\% income-level 2 & 42\% & average speed car -peak hour & 18km/h\\
\% income-level 3 & 24\% & average speed public transit -peak hour & 16km/h\\
\% motorcycle users & 20\% \\
\% car users & 43\% \\
\% public transit users & 37\% \\
accident rate probability for motorcycle & 0.2\\
\hline
\end{tabular}
\end{table}

\section{Accident rate by mode}\label{app:accidents}
\begin{figure}[htb]
{
\centering
\includegraphics[width=0.8\textwidth]{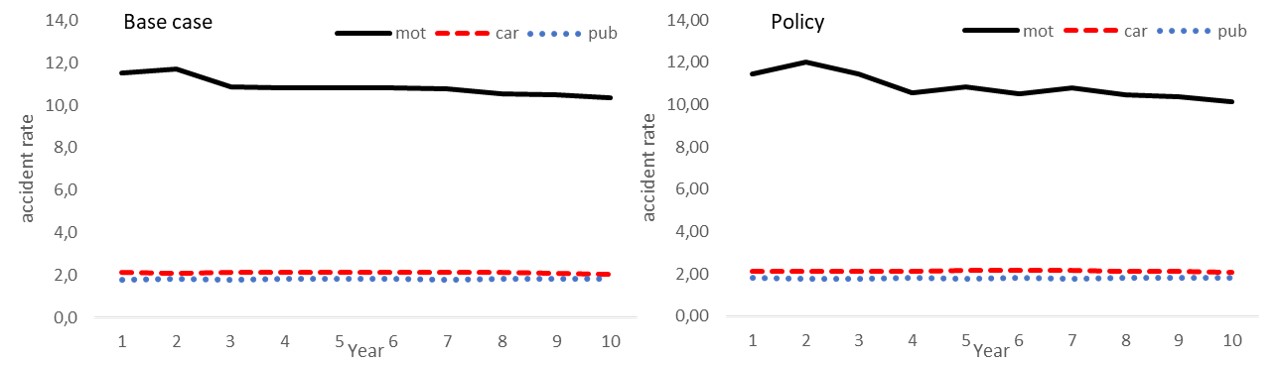}
\caption{Accident rates by mode.}\label{fig:accidents}
}
\end{figure}

Figure \ref{fig:accidents} shows the accident rate for the three transport modes. 

\section{CO2 emissions by mode}\label{app:pollution}
\begin{figure}[htb]
{
\centering
\includegraphics[width=0.8\textwidth]{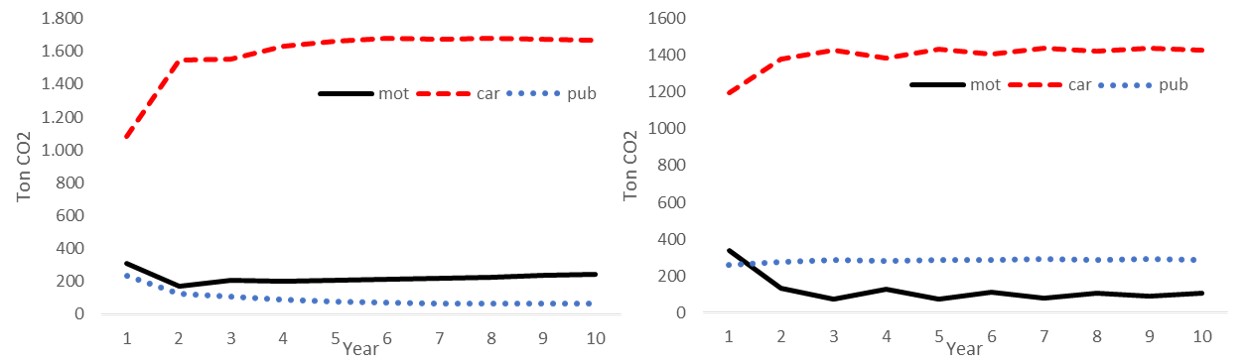}
\caption{CO2 emissions by mode.}\label{fig:emissions}
}
\end{figure}

Figure \ref{fig:emissions} shows the CO2 emissions by mode throughout the simulation. 

\bibliographystyle{scsproc}

\bibliography{Ref}